\documentclass[pdftex,twocolumn,epjc3]{svjour3}          

\RequirePackage[T1]{fontenc}

\smartqed  

\RequirePackage{graphicx}
\RequirePackage{mathptmx}      
\RequirePackage{flushend}
\RequirePackage[numbers,sort&compress]{natbib}
\RequirePackage[colorlinks,citecolor=blue,urlcolor=blue,linkcolor=blue]{hyperref}

\journalname{Eur. Phys. J. C}

\begin{document}

\title{Observation of Time-Dependent Internal Charge Amplification in a Planar Germanium Detector at Cryogenic Temperature }


\author{P. Acharya\thanksref{addr1}
        \and
        M. Fritts\thanksref{addr2} 
        D.-M. Mei\thanksref{e1,addr1}, 
        V. Mandic\thanksref{addr2},
        C.-J. Wang\thanksref{addr1},
        R. Mahapatra\thanksref{addr3} and
        M. Platt\thanksref{addr3}
        }

\thankstext{e1}{Corresponding Author: Dongming.Mei@usd.edu}

\institute{Department of Physics, The University of South Dakota, Vermillion, SD 57069, USA\label{addr1}
\and
School of Physics and Astronomy, University of Minnesota, Minneapolis, MN 55455, USA\label{addr2}
\and
Department of Physics and Astronomy, Texas A$\&$M University, College Station, TX  77843, USA\label{addr3}}
\date{Received: date / Accepted: date}
\maketitle

\begin{abstract}
For the first time, time-dependent internal charge amplification through impact ionization has been observed in a planar germanium (Ge) detector operated at cryogenic temperature. In a time period of 30 and 45 minutes after applying a bias voltage, the charge energy corresponding to a baseline of the 59.54 keV $\gamma$ rays from a $^{241}$Am source is amplified for a short period of time and then decreases back to the baseline. The amplification of charge energy depends strongly on the applied positive bias voltage with drifting holes across the detector. No such phenomenon is visible with drifting electrons across the detector. We find that the observed charge amplification is dictated by the impact ionization of charged states, which has a strong correlation with impurity level and applied electric field. We analyze the dominant physics mechanisms that are responsible for the creation and the impact ionization of charged states. Our analysis suggests that the appropriate level of impurity in a Ge detector can enhance charge yield through the impact ionization of charged states to achieve extremely low-energy detection threshold ($<$ 10 meV) for MeV-scale dark matter searches if the charge amplification can be stabilized.    
 
\end{abstract}
                              
\maketitle
\section{Introduction}
 Dark matter (DM) is believed to be ubiquitous. It makes up 85\% of the mass of the universe~\cite{fzw, ghi, jlf, mwg}. Many candidates including axions, low-mass DM, and weakly interacting massive particles (WIMPs) have been postulated~\cite{axion, lmdm, wimp}. WIMPs are expected to generate observable nuclear recoil energy through elastic scattering off nuclei~\cite{rjg}. Despite great efforts made in searching for axions and WIMPs ~\cite{ardm, cd09, cd, cd1, cd2, cog, cre12, cou, bar, dama, dar, dri, ede, kim, lux, pan, pic, cd14, xe11, xe15, xe17, xma, zep},  DM remains undetected. 
 Recently, low-mass DM in the MeV-scale has become an exciting DM candidate~\cite{ess2012, ess2016, ho, ste}. To directly detect MeV-scale DM, a detector with sensitivity of measuring a single electron-hole (e-h) pair is required, since  the energy deposition induced by MeV-scale DM through elastic scattering off electrons or nuclei is in the range of sub-eV to 100 eV~\cite{ess2012, mei}. In 2018, Mei et al. proposed to detect MeV-scale DM utilizing germanium internal charge amplification (GeICA)~\cite{geia} for the charge created by the ionization of impurities~\cite{mei}. GeICA can potentially achieve a detection energy threshold of $\sim$0.1 eV (100 meV), allowing a large portion of both electronic recoils and nuclear recoils in the range of sub-eV to 100 eV induced by MeV-scale DM to be accessible~\cite{mei}. 
 
 GeICA amplifies internal charge through impact ionization, which is a process first observed in Ge diodes a few decades ago~\cite{imp1, imp}. In this process, a charge carrier, electron or hole, with sufficient kinetic energy can knock a bound electron out of its valence state and elevate it to a state in the conduction band, creating an electron-hole pair.  Carriers gain sufficient kinetic energy through applying a strong electric field. Impact ionization of Ge atoms requires higher electric field than that of impurities, since the bandgap of Ge is about 0.7 eV (700 meV) while the ionization potential of neutral impurity atoms in Ge is around 0.01 eV (10 meV). When a Ge detector is cooled down to below 10 K, the residual impurities in Ge start to freeze out from the conduction or valence band to localized states~\cite{mei2022evidence}. At below 6 K, the localized states become thermally stable and form electric dipole states~\cite{mei2022evidence}, which are excited neutral impurity states with a binding energy of less than 10 meV. The dipole states can trap charge to form cluster dipole states~\cite{mei2022evidence} with even smaller binding energy depending on the operational temperature.

\begin{figure} [htbp]
  \centering
  \includegraphics[clip,width=0.9\linewidth]{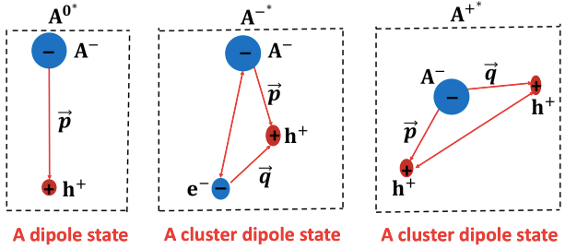}
  \caption{Shown are the processes involved in the formation of the excited dipole states and the cluster dipole states in a p-type Ge detector operated at low temperatures, where $\vec{p}$ and $\vec{q}$ are the corresponding dipole moments.}
  \label{fig:f0}
\end{figure}

The formation of excited dipole states
and cluster dipole states in p-type Ge is depicted in Figure~\ref{fig:f0}~\cite{mei2022evidence}. The phase space of an immobile negative ion for trapping charge carriers is smaller than that of movable bound holes, whose motion is restricted by the Onsager radius, $R=\frac{1}{4\pi\varepsilon\varepsilon_{0} K_{B}T}$, where $\varepsilon$= 16.2 is the relative permittivity for Ge, $\varepsilon_{0}$ is the permittivity of free space, $K_{B}$ is the
Boltzmann constant, and $T$ is temperature. Therefore, the probability of forming $A^{-^{*}}$ states is higher than forming $A^{+^{*}}$ states in a p-type detector. This is why electrons are trapped more severely than holes in a p-type detector.

 \section{The Experimental Methods and the Observed Physical Phenomenon}
 
\begin{figure} [htbp]
  \centering
  \includegraphics[clip,width=0.9\linewidth]{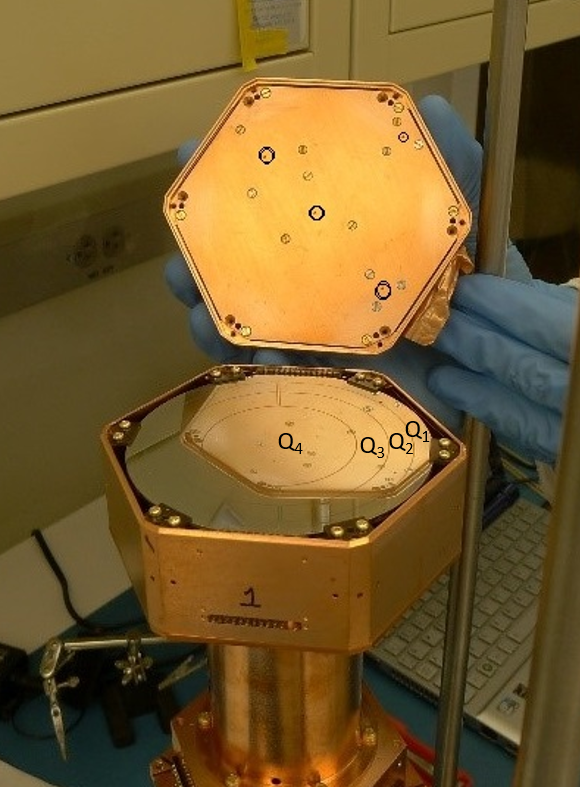}
  \caption{Shown is the detector studied in this work with four electrodes read-out ($Q_{1}, Q_{2}, Q_3, and Q_{4}$). Four $^{241}Am$ sources were arranged in such a way that one source was over each channel. The opposite side of the detector is grounded with uniform Al electrodes so that the electric field points along the z-axis of the crystal. The size of the detector is 10 cm in diameter and 3.3 cm in thickness with a mass of ~$\sim$1.4 kg.}
  \label{fig:f1}
\end{figure}

 The impact ionization of impurities in Ge specimens has been reported for a range of temperatures ($\sim$ 4.2 K to 298 K) by many authors from 1950s - 1970s~\cite{sclar, pickin, smith, palm}. The most recent impact ionization of impurities was reported by Phipps et al. with SuperCDMS-style detectors at 40 milliKelvin (mK)~\cite{phipps1, phipps2}, which is similar to this work with a detector made from a USD crystal~\cite{wang1, wang2}. The detector was fabricated at Texas A $\&$ M University with four channels for charge readout, geometrically similar to SuperCDMS style detectors~\cite{supercdms}, as shown in Figure~\ref{fig:f1}. 
 The detector was wire-bonded, mounted in a dilution refrigerator and tested at the K100 Detector Testing Facility at the University of Minnesota (UMN). 
 
 \begin{figure} [htbp]
  \centering
  \includegraphics[clip,width=0.9\linewidth]{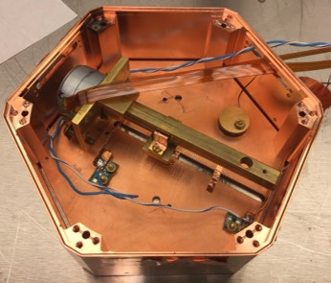}
  \caption{Shown is the orientation of the  Am-241 source mover before it was installed.}
  \label{fig:f11}
\end{figure}
 
The detector was cooled down to $\sim$40 mK and then tested in two separate refrigerator runs, Run~67 in 2018 and Run~74 in 2021. In Run~67 four $^{241}$Am sources were mounted above each channel on the detector (see Figure~\ref{fig:f1}). The lead collimators with 0.2~mm holes allow 59.54 keV $\gamma$ rays through to the detector. Alpha particles from the sources were blocked by the source encapsulation. We observed 59.54 keV peaks in spectra from each channel and performed different measurements over a course of two weeks. In Run~74 a single $^{241}$Am source was mounted on a carriage that could be moved by a superconducting stepper motor (see Figure~\ref{fig:f11}). The source was of a different design with a 0.5~mm collimator hole and a lower rate of $\gamma$ rays incident on the detector (about 75\% as much as the sources used in Run~67).

 \begin{figure}
  \centering
  \includegraphics[clip,width=0.9\linewidth]{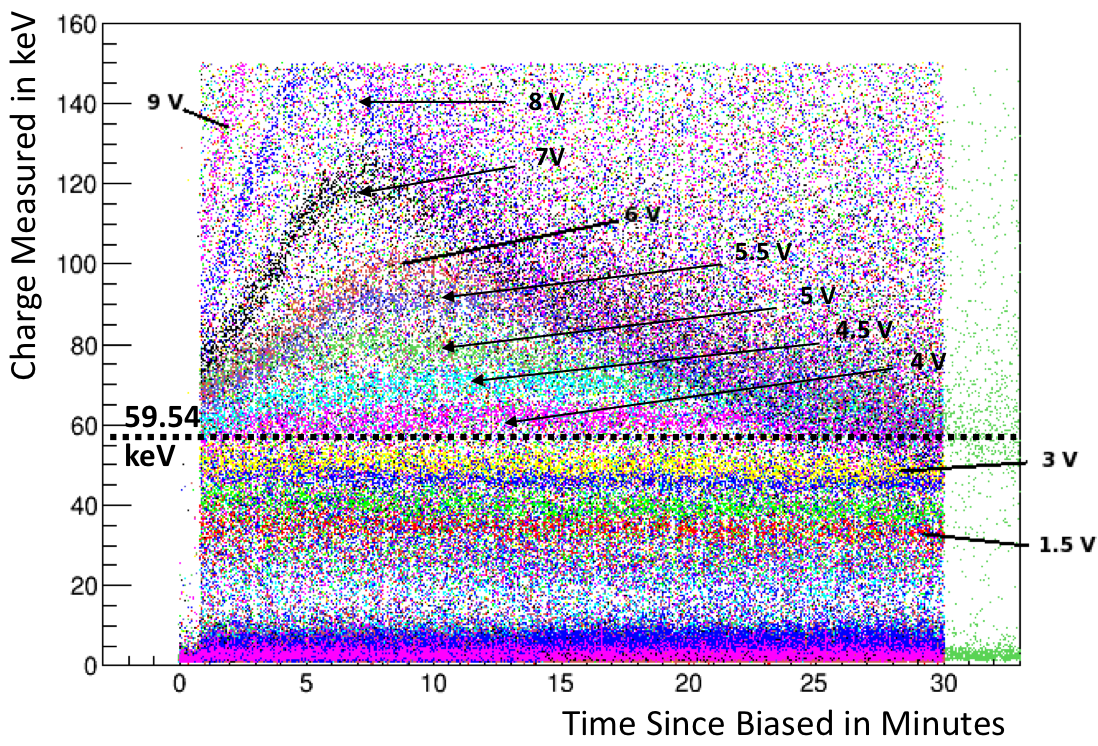}
  \caption{Shown is the time-dependent charge response from Q4 at positive biases for Run~67. The baseline of 59.54 keV gamma ray from Q4 initially rose linearly for a few minutes, then quickly transitioned to an $\sim$exponential fall off over 10s of minutes.}
  \label{fig:f2}
\end{figure}

 \begin{figure}
  \centering
  \includegraphics[clip,width=0.9\linewidth]{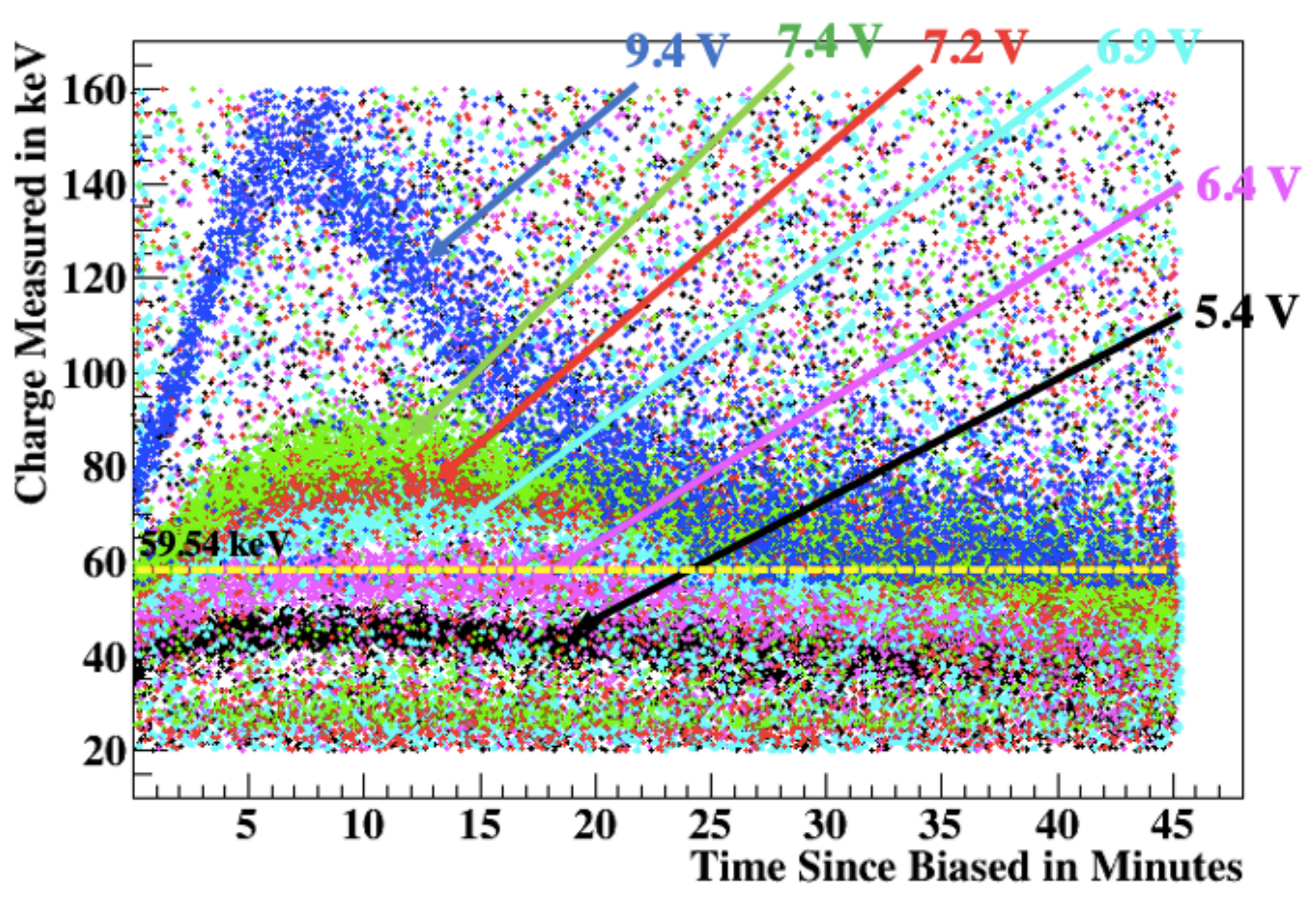}
  \caption{Shown is the time-dependent charge response from Q4 at positive biases for Run~74. The baseline of 59.54 keV gamma ray from Q4 initially rose linearly for a few minutes, then quickly transitioned to an $\sim$exponential fall off over 10s of minutes.}
  \label{fig:f12}
\end{figure}

\section{Experimental Data Analysis and Results}

We used the data measured at various bias voltages to characterize the charge collection efficiency (CCE), defined as the ratio of the measured charge energy to 59.54 keV gamma rays,  as a function of bias for each channel, and studied how the CCE varies with time under a given bias voltage. This work presents the results of an analysis of the 59.54 keV calibration line behavior, as shown in Figure~\ref{fig:f2} \& Figure~\ref{fig:f12}. The observation in particular had a rich set of time dependent behaviors, which are the amplitude of the signals from the 59.54 keV line to initially rise with time after bias, then fall to some steady-state value. This phenomenon was only present with positive biases. Since the $^{241}$Am sources are positioned at the biased side of the crystal and the mean free path of 59.54 keV gamma rays in Ge is about 0.09 cm, this means that this phenomenon is present for events in which electrons are collected almost immediately and holes are drifted through the full detector thickness. This indicates that the observed behavior has a polarity dependence for all events.  

In Run~67 the observed phenomenon was much stronger in the center channel (Q4) and weakest in the outer ring channel (Q1), suggesting a radial dependence. In Run~74, with only one source present with a weaker event rate, the radial dependence is not evident, indicating an overall event rate dependence. In this work, we only study the 59.54 keV line in the center channel (Q4) at positive biases, since this channel outputs the clearest results.

 The observed features of the time-dependent charge collection for the 59.54 keV line can be summarized as: (i) time-dependent impact ionization of drifting holes kicks off at a positive bias of 4.5 volts (1.36 V/cm) and a mixing chamber (MC) temperature in a range from (30-35) mK 
 during  Run~67 and no such a phenomenon is observed with a negatively biased detector when drifting electrons; similarly, we observed the time-dependent impact ionization phenomenon for holes in Run~74 kicking off at a positive bias around 7 volts (2.12 V/cm) and a MC temperature in a range from (35-40) mK (ii) the hole impact ionization initially increases linearly with a rate dependent on the bias and overall event rate; and (iii) after reaching a peak the increased charge signal falls exponentially with a fall time dependent on bias. The loss of both types of signal is easily explained as the loss of CCE due to the breakdown of bulk field from trapped charges. The linear increase of the hole impact ionization signal and lack of electron impact ionization indicate
 that the population of possible hole impact ionization sites starts small and is created by drifting electrons which are captured. This suggests a possible physical mechanism involving the combination of the following three processes: $e^{-} + A^{0^{*}} \rightarrow A^{-^{*}}$, $h^{+} + A^{-^{*}} \rightarrow e^{-} + 2h^{+} + A^{-}$, and $h^{+} + A^{-} \rightarrow A^{0^{*}}$. The first process represents the creation of cluster dipole states ($A^{-^{*}}$) through drifting electrons across the detector induced by background radiation as shown in Figure ~\ref{fig:f0}; the second stands for hole impact ionization of cluster dipole states; and the third is trapping of $h^{+}$, which determines the loss of neutralization at the end of series. 
 
 There is also a possible process, $E_{ph} + A^{-^{*}} \rightarrow e^{-} + A^{0{*}}$, which is the impact ionization of cluster dipole states through absorbing  Neganov-Luke phonons~\cite{luke}, where $E_{ph}$ is the energy of phonons created by drifting holes under a given electric field. This process is not a significant contribution in the experiment since the Neganov-Luke phonons have energy smaller than 1 meV~\cite{mei}. However, the creation of cluster dipole charge states ($e^{-} + A^{0^{*}} \rightarrow A^{-^{*}}$) and the impact ionization of cluster dipole states ($h^{+} + A^{-^{*}} \rightarrow e^{-}$ + 2$h^{+} + A^{-}$) inside the detector generate a dynamic electric field, which, in turn, impacts charge transport and charge creation. As a result, in this model the observed time-dependent impact ionization of 59.54 keV $\gamma$ rays involves the growth of charge states as a function of time, the impact ionization of time-dependent cluster dipole states, and the loss of CCE due to the breakdown of bulk field from trapped charges, which is again a function of time.

 Thus, the analysis presented here attempts to quantify the observed behavior in an empirical way. This will help lend insight into likely physical models that dominate the observed behavior.
 We assume that the detected charge energy, $E(t)$, is related to the input 59.54~keV $\gamma$ rays through the following equation: 
 \begin{equation}
     \label{e1}
     E(t) = E_{\gamma} \{p_{0}+ p_{1}exp[\frac{p_{2}}{p_{3}}(1-exp(-p_{3} t))]\}exp(-p_{4}t),
 \end{equation}
where $E_{\gamma}$ = 59.54 keV. Other terms in Eq.~\ref{e1} are explained below:

   (1) $p_{0} + p_{1}$ represents the average CCE ($\epsilon_{0}$) at $t$ = 0. If it is greater than 1, it means that charge energy is gained through impact ionization. It is expected that the reaction,  $h^{+} + A^{-^{*}} \rightarrow 2h^{+}+ e^{-} + A^{-}$, dominates the hole impact ionization at $t$=0 in the range of the applied field~\cite{sund}.  The parameter $p_{0} + p_{1}$ is the direct measurement of CCE at $t$ = 0 in equation~\ref{e1}, as depicted in Figure~\ref{fig:f2} \& Figure~\ref{fig:f12}.

 (2) $\{p_{0} + p_{1}exp[\frac{p_{2}}{p_{3}}(1-exp(-p_{3} t))]\}exp(-p_{4}t)$ is the average CCE ($\epsilon_{t})$ for the charge created by the impact ionization at $t>0$. 
 We can write $\epsilon_{t}$  = $\epsilon_{0}\times M(t)\times d(t)$, where M(t)  stands for the charge gained through time-dependent impact ionization while $d(t)$ is a time-dependent charge damping factor that describes the charge trapping due to a dynamic process described in points (3) and (4) which creates more charge states. Note that $\epsilon_{0}$, $d(t)$, and $M(t)$ are correlated with applied electric field. 

 (3) $p_{2}$ represents the rate of creating $A^{-^{*}}$ states. It depends on the density of the dipole state, the overall event rate, the drift velocity, and the charge trapping cross section. Since both the drift velocity and the charge trapping cross section are field dependent, it is expected $p_{2}$ has a correlation with applied electric field.  
 
 (4) $p_{3}$ represents the rate of decreasing $A^{-^{*}}$ states, which is proportional to the applied field and the overall event rate.
 
 (5) $exp(-p_{4}t)$ represents the loss of neutralization. If we let $p_{4}=1/\tau$, the factor $exp(-t/\tau)$ describes the collection of charge created by impact ionization falling exponentially as a function of time depending on bias. The parameter $\tau$ measures the effective time constant for the loss of charge signal due to the loss of CCE resulting from the breakdown of bulk field from trapped charges. 
 
\begin{figure} [htp]
 \centering
 \includegraphics[clip,width=0.9\linewidth]{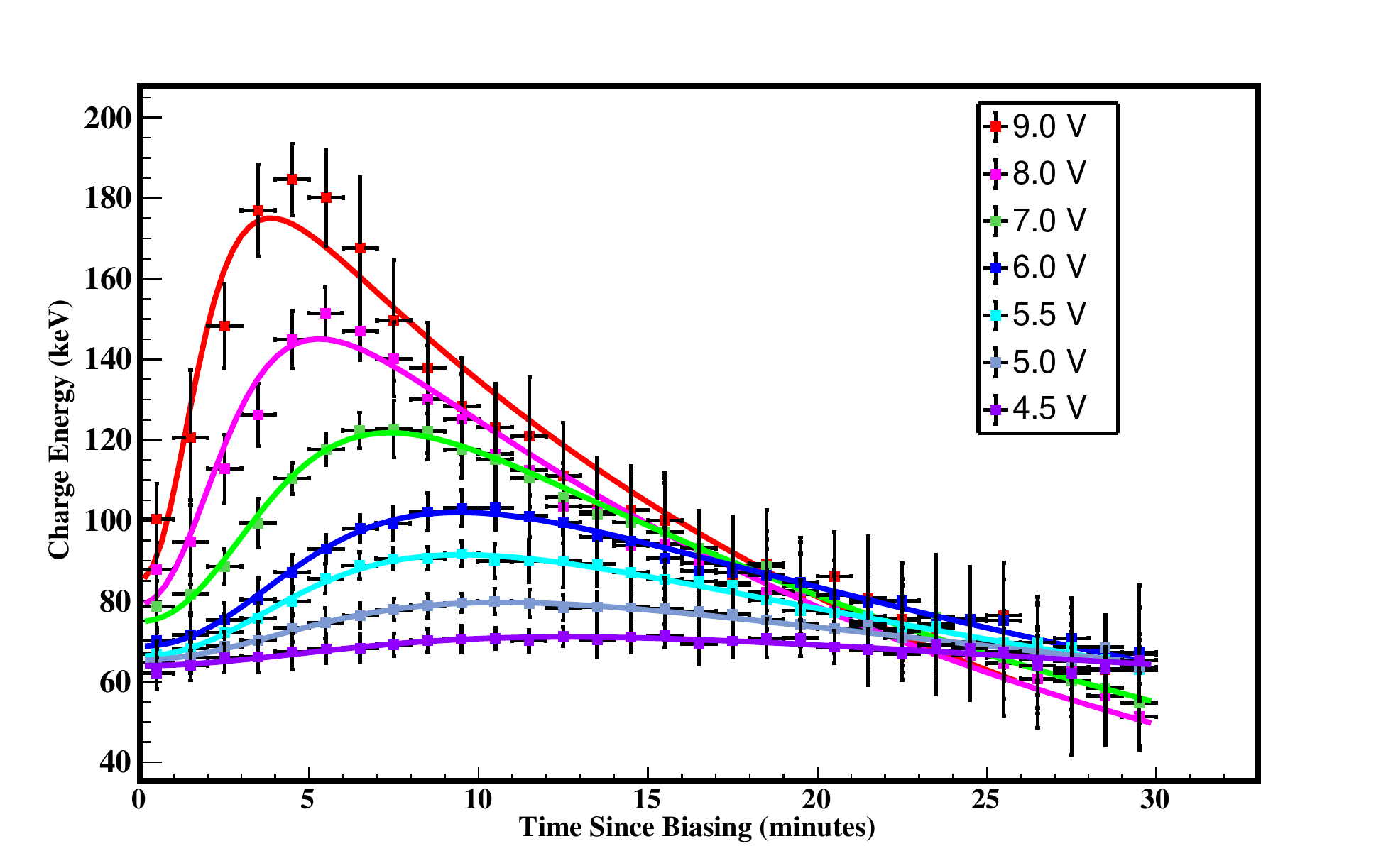}
 \caption{Shown is the detected charge energy created by impact ionization as a function of applied bias voltage fitted by Eq.~\ref{e1} using data from Run~67. The error bars represent the vertical width of the band shown in Figure~\ref{fig:f2}. }
 \label{fig:f4}
\end{figure}

\begin{figure} [htp]
  \centering
  \includegraphics[clip,width=0.9\linewidth]{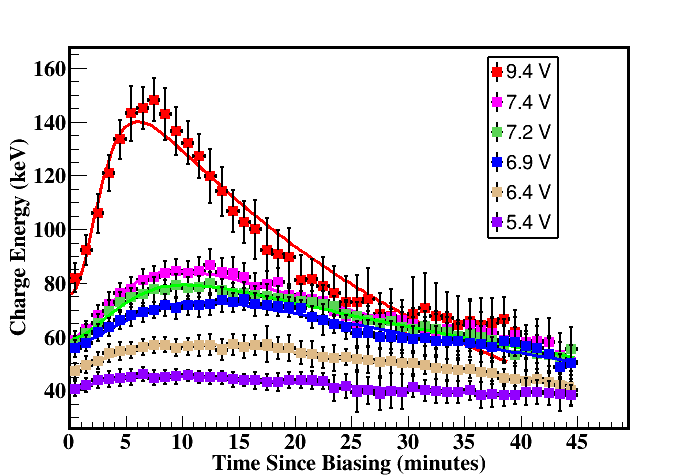}
  \caption{Shown is the detected charge energy created by impact ionization as a function of applied bias voltage from Run~74 fitted by Eq.~\ref{e1}. The error bars represent the vertical width of the band shown in Figure~\ref{fig:f12}. }
  \label{fig:f15}
\end{figure}

 Using this empirical model expressed in Eq.~\ref{e1}, we fit the data from Run~67 and Run~74. The observed trends and the fits are shown for different applied bias voltages in Figures~\ref{fig:f4} and \ref{fig:f15}. 
 Since the two runs were taken at two different periods at which the tower temperature of detector housing was slightly different. Thus, we expect that the detector responds slightly different to the impact ionization as it depends on the detector conditions including the thermal effect. For accurate reading in a temperature, the mixing chamber (MC) temperature was taken. The MC temperature in Run~67 was in a range between (30-35) mK for the data series shown in Figure ~\ref{fig:f4}. Similarly in Run~74, the MC temperature was in a range between (35-40) mK for the data series shown in Figure ~\ref{fig:f15}, being higher compared to Run~67. 
 
 At a given bias the observed phenomenon is stronger in Run~67 than in Run~74. This can be attributed to differences in the overall event rate in the detector between the two runs. By introducing an additional external source it was observed that the impact ionization effect seen in 59.54~keV events was stronger when the overall event rate in the detector was increased. In Run~74 the overall rate was lower because the $^{241}$Am source used produced events at a lower rate, only one source was used rather than four, and a lead shield was erected around the cryostat which reduced the background rate from radioactivity in the lab environment. 
 
\begin{figure} [htp]
  \centering
  \includegraphics[clip,width=0.95\linewidth]{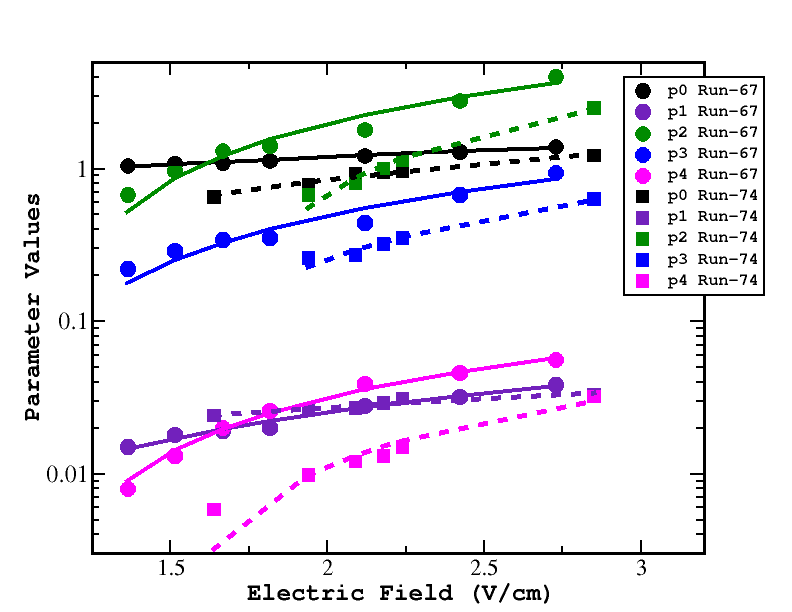}
  \caption{Shown are the fitting parameters, $p_0$ to $p_4$ from equation \ref{e1} as a function of the applied field, $E$, and fits a linear regression model, which demonstrates the correlation of the parameters with the electric field using data from each run. The fitting functions for those parameters in Run~67 (solid lines fitted to the circles) are: $p_{0}$=0.25$E$+0.69, $p_{1}$=0.017$E$-0.0087, $p_{2}$=2.27$E$-2.57, $p_{3}$=0.49$E$-0.49, and $p_{4}$=0.036$E$-0.039. Similarly, the fitting functions for those parameters in Run~74 (dotted lines fitted to the squares ) are: $p_{0}=0.47E-0.11$, $p_{1}=0.0078E+0.011$, $p_{2}=2.1E-3.5$, $p_{3}=0.43E-0.61$, and $p_{4}=0.022E-0.033$.}
  \label{fig:fp}
\end{figure}

We summarize the fitting parameters in Table~\ref{t1}. The fitting to the impact ionization curve using Eq.~\ref{e1} shows that all the parameters have a good correlation with
the electric field.
This indicates that the parameters have their own effect under the electric field. 
The variation of those parameters with the electric field is shown in Figure~\ref{fig:fp}. Since each fitting parameter is a combination of at least two physical processes, it is difficult to figure out the impact of the applied field on any single process. 

\begin{table} [htp]
\centering
\caption{A summary of the fit parameters using equation~\ref{e1}. Note that the values of $p_2$ and $p_3$ are omitted for 5.4 V in Run 74 due to the rate of creating and decreasing $A^{-*}$ states was very small, leading to abnormal values for $p_2$ and $p_3$, which do not follow the trend observed at higher bias voltages, where the dependency on the applied bias field is evident.}
\label{t1}
\begin{tabular}{ |p{0.8cm}|p{0.8cm}|p{0.8cm}||p{0.8cm}||p{0.8cm}||p{0.8cm}| }
\hline
\multicolumn{6}{|c|}       {Run~67} \\
\hline
Bias&$p_{0}$&$p_{1}$&$p_{2}$&$p_{3}$&$p_{4}$\\
\hline
4.5 V&1.05&0.015&0.67&0.22&0.0080\\
\hline
5.0 V&1.07&0.018&0.97&0.29&0.013\\
\hline
5.5 V&1.09&0.019&1.31&0.34&0.020\\
\hline
6.0 V&1.13&0.020&1.41&0.35&0.026\\
\hline
7.0 V&1.22&0.028&1.80&0.44&0.039\\
\hline
8.0 V&1.28&0.032&2.79&0.67&0.046\\
\hline
9.0 V&1.39&0.038&3.98&0.94&0.056\\
\hline
\hline
\multicolumn{6}{|c|} {Run~74} \\
\hline
Bias&$p_{0}$&$p_{1}$&$p_{2}$&$p_{3}$&$p_{4}$\\
\hline
5.4 V&0.65&0.024&-&-&0.0058\\
\hline
6.4 V&0.78&0.026&0.67&0.26&0.0098\\
\hline
6.9 V&0.92&0.027&0.80&0.27&0.012\\
\hline
7.2 V&0.95&0.029&1.002&0.32&0.013\\
\hline
7.4 V&0.96&0.031&1.13&0.35&0.015\\
\hline
9.4 V&1.22&0.033&2.50&0.63&0.032\\
\hline
\end{tabular}
\end{table}

 \section{Discussion and Perspective of the Physics Application}
 
 As can be seen from Figure~\ref{fig:fp}, although the parameters, $p_{0}$, $p_{1}$, $p_{2}$, $p_{3}$, and $p_{4}$ obtained from Run 67 and Run 74 have similar tendencies as a function of applied bias field, the values of these parameters differ between the two runs from a minimum of a few percent to a maximum of $\sim$40\%, depending on the applied bias field. The cause of this difference is likely due to the overall event rates used in the two runs. Run 67 has a higher event rate compared to Run 74.

For bias 4.5~V and above in Run~67, the value of $p_{0}+p_{1}$ is greater than 1.0, indicating impact ionization at $t=0$. In Run~74 the bias required to produce impact ionization at $t=0$ was greater, 9.4~V. The difference between the runs is apparently attributable to the higher event rate in Run~67 mentioned previously.

 For stable operation at which the average CCE is not a function of time for a given electric field, the parameters $p_{3}$ and $p_{4}$ from the empirical model (Eq.\ref{e1}) should be zero. Using the fitted functions obtained from Run 67,  $p_{3}$ = 0 and $p_{4}$ = 0 at electric fields less than $\sim$1~V/cm. At such a small electric field, the measured CCE, $p_{0}$ + $p_{1}$ = 0.9483 and is constant in time. This means that at this electric field, the detector is a normal detector, with no visible impact ionization. The charge collection efficiency is consistent with typical SuperCDMS detectors operated at a similar electric field~\cite{phipps2} and suggests that detectors made from USD-grown crystals are suitable as SuperCDMS-style detectors in terms of charge collection.

\begin{figure} [htp]
  \centering
  \includegraphics[clip,width=0.95\linewidth]{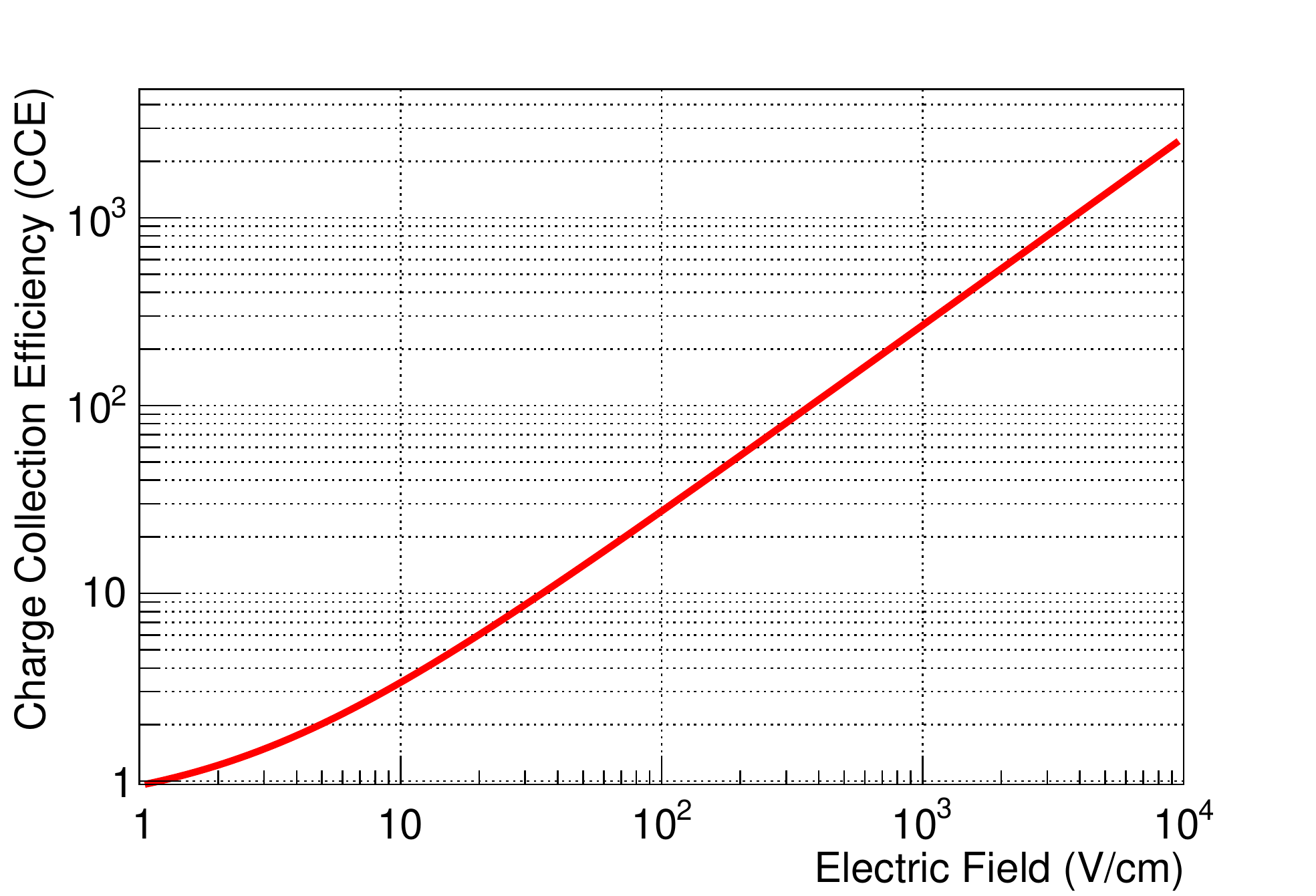}
  \caption{Shown is the predicted CCE as a function of applied electric field. }
  \label{fig:am}
\end{figure}

At a higher electric field, for example at the applied bias of 9 V, the CCE is about 1.4 at $t = 0$ from Run 67. This means a higher electric field is needed to generate impact ionization. If one assumes the charge collection time per event is less than 2 microseconds in this large-size detector, we can predict the CCE as a function of applied electric field using the fitted parameters from Run 67 and the empirical model (Eq.~\ref{e1}). Figure~\ref{fig:am} shows that the CCE can be increased by a factor of $\sim$100 when the detector is operated at a field of $\sim$ 400 V/cm with a charge collection time of 2 microseconds. This means that the charge can be amplified by a factor of $\sim$100 at a bias voltage of $\sim$1300 volts. 
Note that this extrapolation was obtained by extending the fitted empirical trend as a function of applied electric field by more than two orders of magnitude beyond the data. Therefore, the discussion above needs to be verified using experimental data in the future.

 \section{Conclusion} 
 
 We conclude that the observed time-dependent behavior of charge amplification for the 59.54 keV line is mainly due to the impact ionization of charged states by drifting holes across the detector. We attempted to set out possible mechanisms for the impact ionization, which is mainly due to the creation of cluster dipole states, and impact ionization of cluster dipole states. These two processes dominate the production of additional charge that contributes to the amplification of charge while drifting holes across the detector. 
 Our results suggest that a detector with an appropriate density of charged states, created by controlled radiation from the appropriate density of neutral states,  can be developed for searching for MeV-scale dark matter with a detection threshold as low as $<$10 meV, which corresponds to recoil energy induced by sub-MeV dark matter particles. The challenge is to test such a detector at a higher bias voltage such as 1300 volts without triggering avalanche breakdown. The detector avalanche breakdown occurs at about 12 volts in Run 67 and about 15 volts in Run 74. This could be increased by improving the fabrication of the electrical contacts. But the overall event rate, from the $^{241}$Am source plus background radiation, is also relevant to breakdown. Run~74 has a somewhat lower event rate compared to Run~67, and the avalanche breakdown voltage was somewhat higher. We expect the avalanche breakdown voltage can be much higher if events of only a few electron-hole pairs are generated by a well-controlled optical source with the detector being well-shielded from background radiation. More research and development is needed.

 \section{Acknowledgement} 
 This work was supported in part by NSF OISE 1743790, NSF PHYS 1902577, DOE grant DE-FG02-10ER46709, DE-SC0004768, and a research center supported by the State of South Dakota.


\begin{thebibliography}{99}
\bibitem{fzw} F. Zwicky, “Die Rotverschiebung von extragalaktischen Nebeln,” Helv. Phys. Acta 6 (1933) 110-127.
\bibitem{ghi} G. Hinshaw et al., (WMAP Collaboration), “Nine-Year Wilkinson Microwave Anisotropy 
Probe (WMAP) Observations: Cosmological Parameter Results,” ArXiv: 1212.5226, APJS, 208 (2013) 19.
\bibitem{jlf} J. L. Feng, “Supersymmetry and Cosmology,” Annals of Physics 315, Issue 1 (2005) 2-15,
arXiv:hep-ph/0405215.     
\bibitem{mwg} M. W. Goodman and E. Witten, “Detectability of certain dark-matter candidates,” Phys. Rev.
D 31 (June, 1985) 3059-3063.      
\bibitem{axion}  Peccei, Roberto D.; Quinn, Helen R. (20 June 1977). "CP Conservation in the Presence of Pseudoparticles". Physical Review Letters. 38 (25): 1440–1443. Bibcode:1977PhRvL..38.1440P. doi:10.1103/PhysRevLett.38.1440
\bibitem{lmdm} James Bateman, Ian McHardy, Alexander Merle, Tim R. Morris, and Hendrik Ulbricht, "On the Existence of Low-Mass Dark Matter and its Direct Detection," Sci Rep. 2015: 5: 8058. 
\bibitem{wimp} Ungman, Gerard; Kamionkowski, Marc; Griest, Kim (1996). "Supersymmetric dark matter". Physics Reports. 267 (5–6): 195–373. arXiv:hep-ph/9506380.
\bibitem{rjg} R. J. Gaitskell, “Direct Detection of Dark Matter,” Ann. Rev. Nucl. Part. Sci. 54 (2004) 315-
359.
\bibitem{ardm} The ADMX Collaboration; Asztalos, S.J.; Carosi, G.; Hagmann, C.; Kinion, D.; van Bibber, K.; Hotz, M.; Rosenberg, L.; Rybka, G.; Hoskins, J.; Hwang, J.; Sikivie, P.; Tanner, D. B.; Bradley, R.; Clarke, J. (28 January 2010). "A SQUID-based microwave cavity search for dark-matter axions". Physical Review Letters. 104 (4): 041301. 
\bibitem{cd09} Z. Ahmed et al. (CDMS Collaboration), \emph{Search for Weakly Interacting Massive Particles with the First Five-Tower Data from the Cryogenic Dark Matter Search at the Soudan Underground Laboratory}, \emph{Phys. Rev. Lett.} {\bf 102} (2009) 011301.
\bibitem{cd} S.K. Liu et al. (CDEX Collaboration), \emph{Limits on light WIMPs with a germanium detector at 177 eVee threshold at the China Jinping Underground Laboratory}, \emph{Phys. Rev. D} {\bf 90} (2014) 032003.
\bibitem{cd1} H. Jiang  et al. (CDEX Collaboration), \emph{Limits on Light Weakly Interacting Massive Particles from the First 102.8 kg $\times$ day Data of the CDEX-10 Experiment}, \emph{Phys. Rev. Lett.} {\bf 120} (2018) 241301.
\bibitem{cd2} L. T. Yang  et al. (CDEX Collaboration), \emph{Search for Light Weakly-Interacting-Massive-Particle Dark Matter by Annual Modulation Analysis with a Point-Contact Germanium Detector at the China Jinping Underground Laboratory}, \emph{Phys. Rev. Lett.} {\bf 123} (2019), 221301.
\bibitem{cog} Aalseth et al. (CoGeNT Collaboration), \emph{Experimental constraints on a dark matter origin for the DAMA annual modulation effect}, \emph{Phys. Rev. Lett.} {\bf 101} (2008) 251301.
\bibitem{cre12} G. Angloher et al. (CRESST Collaboration), \emph{Results from 730 kg days of the CRESST-II Dark Matter Search}, \emph{Eur. Phys. J. C} {\bf 72} (2012) 1971.
\bibitem{cou} E. Behnke et al. (COUPP Collaboration), \emph{First dark matter search results from a 4-kg CF$_{3}$I bubble chamber operated in a deep underground site}, \emph{Phys. Rev. D} {\bf 86} (2012) 052001. 
\bibitem{bar} J. Barreto et al., \emph{Direct Search for Low Mass Dark Matter Particles with CCDs}, \emph{Phys. Lett. B} {\bf 711} (2012) 264.
\bibitem{dama} R. Bernabei et al. (DAMA/LIBRA Collaboration), \emph{New results from DAMA/LIBRA}, \emph{Eur. Phys. J. C} {\bf 67} (2010) 39.
\bibitem{dar} M. Bossa et al. (DarkSide Collaboration), \emph{DarkSide-50, a background free experiment for dark matter searches}, \emph{JINST} {\bf 9} (2014) C01034.
\bibitem{dri} J.B.R. Battat et al. (DRIFT Collaboration), \emph{First background-free limit from a directional dark matter experiment: results from a fully fiducialised DRIFT detector}, \emph{Phys. Dark Univ.} {\bf 9} (2014) 1.
\bibitem{ede} E. Armengaud et al. (EDELWEISS Collaboration), \emph{Background studies for the EDELWEISS dark matter experiment}, \emph{Astropart. Phys.} {\bf 47} (2013) 1.
\bibitem{kim} S.C. Kim et al. (KIMS Collaboration), \emph{The recent results from KIMS experiment}, \emph{J. Phys. Conf. Ser.} {\bf 384} (2012) 012020.
\bibitem{lux} D. S. Akerib et al. (LUX Collaboration), \emph{Results from a search for dark matter in the complete LUX exposure}, \emph{Phys. Rev. Lett.} {\bf 118} (2017) 021303.
\bibitem{pan} M. Xiao et al. (PandaX Collaboration), \emph{First dark matter search results from the PandaX-I experiment}, \emph{Sci. China Phys. Mech. Astron.} {\bf 57} (2014) 2024.
\bibitem{pic} C. Amole et al. (PICO Collaboration), \emph{Dark Matter Search Results from the PICO-2L C$_{3}$F$_{8}$ Bubble Chamber}, \emph{Phys. Rev. Lett.} {\bf 114} (2015) 231302.
\bibitem{cd14} R. Agnese et al. (SuperCDMS Collaboration), \emph{Search for Low-Mass WIMPs with SuperCDMS}, \emph{Phys. Rev. Lett.} {\bf 112} (2014) 241302.
\bibitem{xe11} E. Aprile et al. (XENON Collaboration), \emph{Design and Performance of the XENON10 Dark Matter Experiment}, \emph{Astropart. Phys.} {\bf 34} (2011) 679. 
\bibitem{xe15} E. Aprile et al. (XENON Collaboration), \emph{Exclusion of leptophilic dark matter models using XENON100 electronic recoil data}, \emph{Science} {\bf 349} (2015) 851.
\bibitem{xe17} E. Aprile et al. (XENON Collaboration), \emph{First Dark Matter Search Results from the XENON1T Experiment}, \emph{Phys. Rev. Lett.} {\bf119} (2017) 181301.
\bibitem{xma} K. Abe et al. (XMASS Collaboration), \emph{Distillation of Liquid Xenon to Remove Krypton}, \emph{Astropart. Phys.} {\bf 31} (2009) 290.
\bibitem{zep} D.Y. Akimov et al. (ZEPLIN-III Collaboration), \emph{Limits on inelastic dark matter from ZEPLIN-III}, \emph{Phys. Lett. B} {\bf 692} (2010) 180.
\bibitem{ess2012} R. Essig, J. Mardon and T. Volansky, \emph{Direct detection of sub-GeV dark matter}, \emph{Phys. Rev. D} {\bf 85} (2012) 076007.
\bibitem{ess2016} R. Essig et al., \emph{Direct detection of sub-GeV dark matter with semiconductor targets},  \emph{J. High Energ. Phys.} {\bf 2016} (2016) 46. 
\bibitem{ho} C.M. Ho and R.J. Scherrer, \emph{Limits on MeV dark matter from the effective number of neutrinos}, \emph{Phys. Rev. D} {\bf 87} (2013) 023505.
\bibitem{ste} G. Steigman, \emph{Equivalent neutrinos, light WIMPs, and the chimera of dark radiation}, \emph{Phys. Rev. D} {\bf 87} (2013) 103517. 
\bibitem{mei} \bibitem{mei} D.-M. Mei et al., \emph{Direct detection of MeV-scale dark matter utilizing germanium internal amplification for the charge created by the ionization of impurities}, \emph{Eur. Phys. J. C} {\bf 78} (2018) 187.
\bibitem{geia} \bibitem{sta} A. S. Starostin and A. G. Beda,  Phys. Atom. Nucl. 63 (2000) 1297-1300, arXiv:hepex/0002063v1.
\bibitem{imp} W. Pickin, "Impact ionization in n-germanium, 4 - 9.5 K," Phys. Rev. B V20 (1979) 2451.
\bibitem{imp1} N. Sclar and E. Burstein, "Impact ionization of impurity in germanium," J. of Phys. and Chem. of Solids, V2 (1957) 1-23. 
\bibitem{mei2022evidence} Mei, D-M and Panth, R and Kooi, K and Mei, H and Bhattarai, S and Raut, M and Acharya, P and Wang, G-J, "Evidence of cluster dipole states in germanium detectors operating at temperatures below 10 K",  AIP Advances 12, 065113 (2022).
\bibitem{sclar} N. Sclar and E. Burstein, "Impact Ionization of Impuriteis in Germanium," J. Phys. Chem. Solids. Pergamon Press 1957. Vol. 2 pp. 1-23. 
\bibitem{pickin} W. Pickin, "Impact ionization in n-germanium, 4.5 - 9.5 K," Phys. Rev. B, V 20 (1979) 2451. 
\bibitem{smith} D. L. Smith, D. S. Pan, and T. C. McGill, "Impact ionization of excitons in Ge and Si," Phys. Rev. B, V 12 (1975) 4360.
\bibitem{palm} J. F. Palmier, "Impact ionization in n-type germanium," Phys. Rev. B, V 6 (1972) 4557. 
\bibitem{phipps1} A. Phipps, B. Sadoulet, K. M. Sundqvist, "Observation of Impact Ionization of Shallow States in Sub-Kevin, High-Purity Germanium," J. Low Temp. Phys. (2016) 184: 336 - 343. 
\bibitem{phipps2} Arran Thomas James Phipps, "Ionization Collection in Detectors of the Cryogenic Dark Matter Searches," PhD Thesis, University of California, Berkeley, Spring of 2016. 
\bibitem{wang1} G.-J. Wang et al., \emph{Development of large size high-purity germanium crystal growth}, \emph{Journal of Crystal Growth} {\bf 352} (2012) 27-30.
\bibitem{wang2} G.-J. Wang et al., \emph{High purity germanium crystal growth at the University of South Dakota}, \emph{J. Phys.: Conf. Ser.} {\bf 606} (2015) 012012.
\bibitem{supercdms} SuperCDMS Collaboration, R. Agnese et al.,  Phys. Rev. Lett. 112 (2014) 241302, arXiv:1402.7137.
\bibitem{luke} P. Luke, "Voltage-Assisted Calorimetric Ionization Detector," Journal of Applied Phsyics 64, 6858-6860 (1988).
\bibitem{sund} K. Sundqvist, "Carrier Transport  and Related Effects in Detectors of Cryogenic Dark Matter Searches," PhD thesis, University of California, Berkeley, 2012. 


\end{thebibliography}
\end{document}